\newif\ifsubmode
\submodetrue

\ifsubmode
  \documentclass[12pt,preprint]{aastex}
  \revised{}
  \accepted{}
\else
  \documentclass[12pt,preprint]{emulateapj}

\fi

\received{2003 May 23}
\begin{document}
\def\hh{\, h^{-1}}
\newcommand{\wth}{$w(\theta)$}
\newcommand{\xir}{$\xi(r)$}
\newcommand{\Lya}{Ly$\alpha$}
\newcommand{\Lyb}{Lyman~$\beta$}
\newcommand{\Hb}{H$\beta$}
\newcommand{\msun}{M$_{\odot}$}
\newcommand{\sfr}{M$_{\odot}$ yr$^{-1}$}
\newcommand{\dnsty}{$h^{-3}$Mpc$^3$}
\newcommand{\za}{$z_{\rm abs}$}
\newcommand{\ze}{$z_{\rm em}$}
\newcommand{\cmtwo}{cm$^{-2}$}
\newcommand{\nhi}{$N$(H$^0$)}
\newcommand{\degpoint}{\mbox{$^\circ\mskip-7.0mu.\,$}}
\newcommand{\halpha}{\mbox{H$\alpha$}}
\newcommand{\hbeta}{\mbox{H$\beta$}}
\newcommand{\hgamma}{\mbox{H$\gamma$}}
\newcommand{\minpoint}{\mbox{$'\mskip-4.7mu.\mskip0.8mu$}}
\newcommand{\mv}{\mbox{$m_{_V}$}}
\newcommand{\Mv}{\mbox{$M_{_V}$}}
\newcommand{\peryr}{\mbox{$\>\rm yr^{-1}$}}
\newcommand{\secpoint}{\mbox{$''\mskip-7.6mu.\,$}}
\newcommand{\sqdeg}{\mbox{${\rm deg}^2$}}
\newcommand{\squig}{\sim\!\!}
\newcommand{\subsun}{\mbox{$_{\twelvesy\odot}$}}
\newcommand{\et}{{\it et al.}~}
\newcommand{\er}[2]{$_{-#1}^{+#2}$}
\def\h50{\, h_{50}^{-1}}
\def\hbl{km~s$^{-1}$~Mpc$^{-1}$}
\def\ltsima{$\; \buildrel < \over \sim \;$}
\def\simlt{\lower.5ex\hbox{\ltsima}}
\def\gtsima{$\; \buildrel > \over \sim \;$}
\def\simgt{\lower.5ex\hbox{\gtsima}}
\def\arcs{$''~$}
\def\arcm{$'~$}
\newcommand{\wu}{$U_{300}$}
\newcommand{\wb}{$B_{435}$}
\newcommand{\wv}{$V_{606}$}
\newcommand{\wi}{$i_{775}$}
\newcommand{\wz}{$z_{850}$}
\newcommand{\hmpc}{$h^{-1}$Mpc}
\newcommand{\ndat}{\nodata}
\newcommand{\hst}{{\it HST}}
\newcommand{\cxo}{{\it CXO}}


\newcommand{\figonecap}{{Exposure map of the GOODS HDF-N Observations. In this
image, blue represents the {\it Chandra} (\cxo) exposure map for the 2
Msec observations described by Alexander et al.\ (2003). Green represents the
current {\it HST}/ACS exposure map, and red represents the planned {\it
SIRTF}/IRAC exposure map. Where all fields overlap, the colors sum to give
white in the representation.  The different ACS tiling patterns on even and
odd epochs produce the sawtooth pattern around the edge of the ACS fields. A
guide-star failure on one of the ACS tiles produces the reddish square seen
within the ACS rectangle. This exposure will be repeated to even out the final
exposure map.  The two planned 75-hour IRAC ultradeep fields can be seen
faintly within the ACS area. These ultradeep fields are planned only for the
HDF-N.  }}

\newcommand{\figtwocap}{{Exposure map of the GOODS CDF-S Observations.
The color-map is the same as for Fig. 1. The {\it Chandra} observations
are those of Giacconi et al.\ (2002).}}

\newcommand{\figthreecap}{{Schematic outline of the GOODS data sets in the CDF-S
region. The blue shows the outline of the ACS area.  The other colors
outline the areas covered by the ESO data described in the text. 
}}

\newcommand{\figfourcap}{{Completeness limits for the GOODS ACS
$z$-band catalog as a function of galaxy total magnitude and
half-light radius. The contours show the percentage of galaxies
recovered by SExtractor in the simulations described in the text. 
}}

\title{The Great Observatories Origins Deep Survey: 
Initial Results From Optical and Near-Infrared Imaging
\altaffilmark{1}}

\author{\sc M. Giavalisco\altaffilmark{2}, 
H. C. Ferguson\altaffilmark{2,12}, 
A. M. Koekemoer\altaffilmark{2},
M. Dickinson\altaffilmark{2,12}, 
D. M. Alexander \altaffilmark{3}, 
F. E. Bauer \altaffilmark{3}, 
J. Bergeron\altaffilmark{4}, 
C. Biagetti\altaffilmark{2},
W. N. Brandt\altaffilmark{3}, 
S. Casertano\altaffilmark{2},
C. Cesarsky\altaffilmark{5}, 
E. Chatzichristou\altaffilmark{6},
C. Conselice\altaffilmark{7}, 
S. Cristiani\altaffilmark{8}, 
L. Da Costa\altaffilmark{5}, 
T. Dahlen\altaffilmark{2}, 
D. De Mello\altaffilmark{9},
P. Eisenhardt\altaffilmark{10}, 
T. Erben\altaffilmark{19},
S. M. Fall\altaffilmark{2}, 
C. Fassnacht\altaffilmark{11}, 
R. Fosbury\altaffilmark{14,17}, 
A. Fruchter\altaffilmark{2}, 
Jonathan. P. Gardner\altaffilmark{9}, 
N. Grogin\altaffilmark{12}, 
R. N. Hook\altaffilmark{5,17}, 
A. E. Hornschemeier\altaffilmark{12},
R. Idzi\altaffilmark{12}, 
S. Jogee\altaffilmark{2}, 
C. Kretchmer\altaffilmark{12},
V. Laidler\altaffilmark{2}, 
K. S. Lee\altaffilmark{12}, 
M. Livio\altaffilmark{2}, 
R. Lucas\altaffilmark{2}, 
P. Madau\altaffilmark{13},
B. Mobasher\altaffilmark{2,14}, 
L. A. Moustakas \altaffilmark{2},
M. Nonino\altaffilmark{8}, 
P. Padovani\altaffilmark{2,14}, 
C. Papovich\altaffilmark{15}, 
Y. Park\altaffilmark{12},
S. Ravindranath\altaffilmark{2}, 
A. Renzini\altaffilmark{5},
M. Richardson\altaffilmark{2}, 
A. Riess\altaffilmark{2},
P. Rosati\altaffilmark{5}, 
M. Schirmer\altaffilmark{18,19},
E. Schreier\altaffilmark{2},
R. S. Somerville\altaffilmark{2},
H. Spinrad\altaffilmark{16}, 
D. Stern\altaffilmark{10},
M. Stiavelli\altaffilmark{2}, 
L. Strolger\altaffilmark{2},
C. M. Urry\altaffilmark{6}, 
B. Vandame\altaffilmark{5},
R. Williams\altaffilmark{2},
C. Wolf\altaffilmark{20}
}

\affil{$^{2}$Space Telescope Science Institute, 3700 San Martin Dr.,
       Baltimore, MD 21218} 
\affil{$^{3}$Department of Astronomy and Astrophysics, Pennsylvania State
       University, 525 Davey Lab, State College, PA 16802}
\affil{$^{4}$Institut d'Astrophysique de Paris -- CNRS, 98bis Boulevard Arago, 
       F--75014, Paris, France}
\affil{$^{5}$European Southern Observatory, Karl--Schwarzschild--Strasse 2,
       D--85748 Garching bei M\"unchen, Germany}
\affil{$^{6}$Department of Astronomy, Yale University, PO Box 208101, New
       Haven, CT 06520}
\affil{$^{7}$Palomar Observatory, California Institute of Technology, Mail Stop
       105-24, Pasadena, CA 91125}
\affil{$^{8}$Istituto, Nazionale di Astrofisica, Osservatorio Astronomico di
       Trieste, via G.B. Tiepolo 11, Trieste, I--34131, Italy}
\affil{$^{9}$NASA Goddard Space Flight Center, Laboratory for Astronomy and
       Solar Physics, Code 681, Greenbelt, MD 20771} 
\affil{$^{10}$Jet Propulsion Laboratory, California Institute of Technology,
       Mail Stop 169-506, Pasadena, CA 91109}
\affil{$^{11}$Department of Physics, University of California, Davis, 1 Shields
       Ave. Davis, CA 95616}
\affil{$^{12}$Department of Physics and Astronomy, The Johns Hopkins
       University, 3400 N. Charles St., Baltimore, MD 21218--2686}
\affil{$^{13}$Department of Astronomy and Astrophysics, University of
       California, Santa Cruz, 1156 High Street, Santa Cruz, CA 95064}
\affil{$^{14}$European Space Agency, Space Telescope Division}
\affil{$^{15}$Steward Observatory, University of Arizona, 933 Cherry Ave.,
       Tucson, AZ 85721--0065}
\affil{$^{16}$Department of Astronomy, University of California, Berkeley, Mail
       Code 3411, Berkeley, CA 94720}
\affil{$^{17}$Space Telescope European Coordinating Facility, 
       European Southern Observatory, Karl--Schwarzschild--Strasse 2, 
       D--85748 Garching bei M\"unchen, Germany }
\affil{$^{18}$Institut f\"ur Astrophysik und Extraterrestrische Forschung, 
       Universit\"at Bonn, Auf dem H\"ugel 71, Bonn, Germany} 
\affil{$^{19}$Max-Planck-Institut f\"ur Astrophysik, 
       Karl--Schwarzschild--Strasse 1,
       D--85748 Garching bei M\"unchen, Germany}
\affil{$^{20}$Department of Physics, University of Oxford, Keble Road,
       Oxcford, OX1 3RH}

\altaffiltext{1}{Based on observations obtained with the NASA/ESA {\it
Hubble Space Telescope} obtained at the Space Telescope Science Institute,
which is operated by the Association of Universities for Research in
Astronomy, Inc. (AURA) under NASA contract NAS 5-26555; observations collected
at the European Southern Observatory, Chile (ESO Programmes 168.A-0845,
64.O-0643, 66.A-0572, 68.A-0544, 164-O-0561, 169.A-0725, 267.A-5729,
66.A-0451, 68.A-0375, 168.A-0485, 164.O-0561, 267.A-5729, 169.A-0725,
64.O-0621, 170.A-0788); observations collected at the Kitt Peak National
Observatory, National Optical Astronomical Observatories, which is operated by
AURA under cooperative agreement with the National Science Fundation.}

\begin{abstract}

This Special Issue of the Astrophysical Journal Letters is dedicated to
presenting initial results from the Great Observatories Origins Deep Survey
(GOODS) that are primarily, but not exclusively, based on multi--band imaging
data obtained with the {\it Hubble Space Telescope} ({\it HST}) and the
Advanced Camera for Surveys (ACS).  The survey covers roughly 320 square
arcminutes in the ACS F435W, F606W, F814W, and F850LP bands, divided into two
well-studied fields.  Existing deep observations from the {\it Chandra} X-ray
Observatory (\cxo) and ground-based facilities are supplemented with new,
deep imaging in the optical and near-infrared from the European Southern
Observatory (ESO) and from the Kitt Peak National Observatory (KPNO). Deep
observations with the {\it Space Infrared Telescope Facility} (SIRTF) are
scheduled. Reduced data from all facilities are being released worldwide
within three to six months of acquisition.  Together, this data set provides
two deep reference fields for studies of distant normal and active galaxies,
supernovae, and faint stars in our own galaxy. This paper serves to outline
the survey strategy and describe the specific data that have been used in the
accompanying letters, summarizing the reduction procedures and sensitivity
limits.

\end{abstract}


\section{INTRODUCTION}

Observations of representative fields at high galactic latitude have long been
an important tool in our quest to understand the distant universe. The Hubble
Deep Field project (Williams et al.\ 1996; 2000) demonstrated the value of
deep, multicolor \hst\ imaging for studies of galaxy evolution, and the
extraordinary multiplier of this value gained by rapid public dissemination of
the data and coordination of the best observations at all wavelengths on
common survey fields (see Ferguson, Dickinson \& Williams 2000 for a review).
The Great Observatories Origins Deep Survey is the next generation of such
deep surveys, uniting some of the deepest observations from space and
ground-based facilities on common areas of the sky.  The two GOODS fields, the
Hubble Deep Field North (HDF-N) and the Chandra Deep Field South (CDF-S), are
the most data-rich deep survey areas on the sky.  The deepest X--ray
observations from the \cxo\ and {\it XMM--Newton} telescopes have been
obtained at these locations, and deep radio maps are available or are now
being collected.  Observations with {\it SIRTF} were selected as one of the
Legacy projects to be executed during the first year of the mission.  The {\it
SIRTF} Legacy project includes major commitments of observing time from the
National Optical Astronomical Observatories (NOAO) and from the European
Southern Observatory to provide spectroscopy (redshifts) for sources in the
fields to the practical limits of existing instruments and to provide
complementary imaging at near ultraviolet ($U$--band) and near infrared
wavelengths.  The GOODS project was subsequently awarded 398 orbits of time to
observe the two fields with {\it HST} and ACS.

The GOODS project has been designed from the outset as a resource for the
entire astronomical community. Data from the {\it HST} observations are
available from the archive within a few days of the observations.  Reduced
data, with successive degrees of refinement, are available within a few months
of the observations.  A similar strategy is followed for the ESO observations,
and will be followed for {\it SIRTF}.  The publicly released {\it HST} images
have already been exploited by Stanway et al. (2003) and Bunker et al.\ (2003)
to study galaxies at $z > 5$, and by Dawson et al.\ (2002) to study a hard
X-ray emitting spiral galaxy. The ESO near-infrared images have been used by
Roche et al.\ (2003) and Yan et al. (2003).

The GOODS field centers (J2000.0) are $12^{\rm h} 36^{\rm m} 55^{\rm s}, $
$+62^\circ 14^{\rm m} 15^{\rm s}, $ for the HDF-N, and $3^{\rm h} 32^{\rm m}
30^{\rm s}, $ $-27^\circ 48^{\rm m} 20^{\rm s}, $ for the CDF-S. Each field
provides an area of approximately $10\arcmin \times 16 \arcmin$ that will be
in common to all the imaging observations. As of May 2003 the \cxo\ data
are all reduced and available, all of the {\it HST} data have been obtained
and are available, including a best--effort version of the reduced data, while
the recalibrated, mosaiced stacks of the ACS images will be released at the
end of August. Much of the planned optical imaging data from KPNO and ESO have
been obtained and are available. Deep near-IR imaging campaigns with the ESO
Very Large Telescope (VLT) ISAAC camera and the KPNO FLAMINGOs camera are
partially complete, but will require one or two more observing seasons for
complete field coverage. Reduced data from first season (2001) of ISAAC
observing are available. The planned spectroscopic campaign from ESO was begun
in the Fall of 2002 with Focal Reducer/low-dispersion spectrograph (FORS2)
observations of preferrentially red objects.  Observations with the VIMOS
spectrograph are planned for the 2003-2004 CDF-S observing season. The HDF-N
already benefits from deep spectroscopic observations from Cohen et al.\
2000. Several teams are carrying out spectroscopic observations of the HDF-N
from the W.M. Keck Observatory in early 2003, and public releases from a few
of these teams are planned for the second half of this year.

The GOODS ACS observations were spaced in time to permit a search for type Ia
supernovae at redshifts $z \approx 1.4$, and a total of 43 supernovae have
been detected in eight GOODS search epochs. In their letter, Riess et
al. provide details of how the type Ia candidates at $z>1$ are selected for
followup observations from among these sources. The other accompanying letters
papers in this issue present initial findings on galaxies and AGN at moderate
to very high redshifts from the analysis of the available GOODS data. A letter
by Somerville et al.\ presents computations of the expected ``cosmic
variance'' for various source populations based on halo statistics in
Cold-Dark Matter models consistent with the Wilkinson Microwave Anistropy
Probe (WMAP) cosmological parameters (Spergel et al.\ 2003).

\section{The Data Set}

The data set discussed here includes ACS images, deep optical and
near--infrared imaging from KPNO and ESO, and the \cxo\ images. The main
features of each data set are summarized in Table 1, while Table 2 lists the
sensitivity achieved in each wavelength band. Figures 1 and 2 show the
exposure maps of the three space--based data sets overlaid on top of each
other to illustrate the extent of the area in common. Figure 3 shows a
schematic of the layout of the existing imaging coverage in the CDF--S.

\section{The HST ACS Data}

The GOODS/ACS observations consist of imaging in the F435W, F606W, F775W and
F850LP passbands, hereafter referred to as \wb, \wv, \wi\ and \wz. While the
\wb--band images were all acquired at the beginning of the survey, the \wv, 
\wi, and \wz--band images have been carried out in five epochs, separated by
40 to 50 days to optimize the search for high-redshift supernovae. Only 60\%
of the total exposure time in each field (3 epochs out of 5) in the \wv, \wi,
and \wz are presented here, since the remainder of the data were not available
at the time these papers were being prepared. In the odd--numbered epochs
each $10\arcmin \times 16\arcmin$ field is tiled by a grid of $3 \times 5$
individual ACS pointings, including some area of tile overlapping to check
photometric and astrometric consistency. In even-numbered epochs, due to {\it
HST} pointing constraints, the field is rotated by $45^\circ$ and a different
overlapping tiling pattern is used, consisting of 16 separate pointings (see
Figures 1 and 2). The complete set of \wb\ observations were obtained in the
first epoch alone, using the $3 \times 5$ grid, with six exposures per
position.

The exposure times at each epoch are typically 1050, 1050 and 2100 seconds in
the \wv, \wi\ and \wz\ bands, respectively. These are subdivided into 2
exposures in each of the \wv\ and \wi\ bands, and into 4 exposures in the \wz\
band. This strategy ensures good rejection of cosmic ray events in the
single--epoch \wz\ bands, which is where the detection of transients is
carried out. Some cosmic rays remain in the \wv\ and \wi\ single--epoch
images, which can nevertheless generally provide colors of the transients.  In
each case, the telescope field of view is shifted (``dithered'') by a small
amount between individual exposures to allow optimal sampling of the
point--spread function and to remove detector gaps and artifacts.  The
multiple epochs are later combined into a single mosaic, allowing cosmic ray
rejection from at least six images per band. Except for the \wb\ band, the
mosaics used for the current papers were constructed from only three epochs of
observations, and the total exposure times are approximately 7200, 3040, 3040,
and 6280 seconds in the \wb, \wv, \wi, and \wz\ bands, respectively.

Basic reduction of the raw data is carried out through {\tt calacs}, the ACS
calibration pipeline, which applies the basic calibration steps of bias
subtraction, gain correction, and flat-fielding (Pavlovsky et al.\ 2002).  The
data received from the {\it HST} pipeline are then further processed by the
GOODS pipeline using the {\tt multidrizzle} script (Koekemoer et al.\ 2003),
to create a set of geometrically rectified, cosmic-ray cleaned images (within
a given epoch).  These processed ``tiles'' are released as the ``best-effort''
GOODS public release v0.5 via the Multimission Archive at Space Telescope
(MAST).

The early version of the geometric distortion model used in the basic
processing described above proved to have significant problems matching
sources in overlapping tiles.  To rectify this, we derived an astrometric
solution for each tile and each epoch based on sources matched in GOODS
groundbased images.  The reference for the CDF-S image was the $R$--band image
of the field obtained as part of the ESO Imaging Survey (EIS) with the Wide
Field Imager (WFI), which in turn was astrometrically calibrated to the Guide
Star Catalog 2 (GSC-2) and put on the ICRS reference frame.  For HDF--N, the
reference image was an $R$-band Subaru image of the field (Capak et al.\
2003).

The astrometric solution for the $z$-band mosaics has been derived by a
least-squares optimization of the position, orientation, x and y pixel scales,
and axis skew of each tile and epoch, minimizing the inter-epoch variations of
the estimated position for $\sim 2000$ sources.  The estimated rms in the
source position, internal to the solution, is about 0.1 to 0.2 WFC pixels.
For the other ACS bands, the astrometric solution is based on a tile-by-tile
match to $z$-band source positions.  These solutions had a residual rms of
about 0.3 pixels, with larger deviations in some local regions across the
field.  All the exposures were then drizzled (Fruchter \& Hook 2002) onto a
series of images with a common pixel grid, in order to create a clean median
image, which was subsequently used to create a cosmic ray mask for each
exposure. Finally, the individual exposures were drizzled, using the new
masks, onto a final single mosaic for each band, measuring 18000 by 24000
pixels with a scale of 0\farcs05/pixel.

For the purposes of the current papers the astrometric solution appears to be
acceptable, although its deficiencies become apparent for bright stars, where
the cosmic ray rejection algorithm rejects some good pixels due to the slight
misregistrations.  This can bias fluxes (faintward) for the brighter point
sources in the $V_{606}$ and $i_{775}$ images (only). We have verified that
there is little or no photometric effect for even slightly extended sources.
A photometric comparison of the \wv\ magnitudes (SExtractor \verb!MAG_AUTO!
values) in the original WFPC-2 HDF-N and the current mosaic reveals systematic
biases of less than 0.01 magnitudes for galaxies in the range $23<$\wv$ < 26$,
with an rms scatter less than 0.2 mag.

The resulting mosaics have a few remaining blemishes and some misregistrations
between bands at the level of a fraction of a pixel. Slight sky level
variations are also apparent. The masking of satellite trails, and reflection
ghosts is not yet perfect, with residuals apparent in few places. 
The samples of galaxies used for the papers in this volume have all been 
individually inspected to remove objects that are thought to be artifacts.
For most samples such contamination is negligible, but the contamination
is significant for the single-band detections used to search for
galaxies at $z \sim 6$, as discussed in detail by Dickinson et al.\ (2003).

\section{The \cxo\ Data}

{\it Chandra} data in the GOODS fields were taken before the GOODS project
began and are available from the \cxo\ archive (Giacconi et al.\ 2002;
Alexander et al.\ 2003).  The integration times of the {\it Chandra} CDF--S
and HDF--N observations total roughly 1 and 2 Msec, respectively.  The GOODS
{\it SIRTF} field layout was designed in part to maximize the coverage of the
X--ray area.  Subsequently, the ACS fields were designed to optimally cover
both the {\it SIRTF} and {\it Chandra} fields.  The \cxo\ data for both fields
have been re--analyzed in a self--consistent way, and new catalogs have been
published by Alexander et al.\ (2003). Finally, these X--ray catalogs have
been matched to the ACS catalogs by Koekemoer et al.\ (in preparation) for
CDF--S and by Bauer et al.\ (in preparation) for HDF--N.  Over 90\% of the
area covered by the GOODS ACS observations, the $S/N \geq 10$ point source
sensitivity limits in the 0.5--2.0~keV and 2-8~keV bands, respectively, are
$3.3$ and $24\times 10^{-16}$~erg~s$^{-1}$~cm$^{-2}$ for the HDF--N and $4.3$
and $30\times 10^{-16}$~erg~s$^{-1}$~cm$^{-2}$ for the CDF--S.  The
sensitivities at the \cxo\ aim point, where the exposure time and image
quality are maximized, are as much as a factor of $\sim 2$ and $\sim 4$
fainter for the CDF-S and HDF-N, respectively.

\section{Ground--based Near--UV, Optical, and Near--IR Imaging}

Ground--based imaging spanning a wide wavelength baseline is an important
component of GOODS, and here, we describe the data that have been used for
projects presented in this Special Issue. The CDF--S imaging consists of a
complex arrangement of data sets covering different areas, often using
multiple instrument pointings, as illustrated in Figure~3.  These data sets
were all placed on a common astrometric grid tied to the GSC2.  The PSFs were
matched by Gaussian convolution to a common FWHM of 0\farcs9.  The WFI $U'UBV$
images and a few of the images obtained with the New Technology Telescope
(NTT) and the Superb--Seeing Imager (SOFI) have slightly poorer seeing,
0\farcs9--1\farcs05, but most projects use colors measured through $3\arcsec$
apertures, so this small mismatch should have little effect.  The PSF--matched
images were mosaiced onto a common $0\farcs3$ pixel grid with the SWarp
software (Bertin 2002).  Noise maps were constructed for use in object
detection and cataloging, carefully accounting for the effects of interpixel
correlations.  High--resolution ($0\farcs15$/pixel) ISAAC mosaics were also
made without degrading their excellent image quality.  We have used the colors
of stars to check the photometric zeropoints of the CDF--S data, as described
below.  Where possible, these were identified using the ACS imaging, using
criteria for star--galaxy separation based on a diagram of peak surface
brightness vs.\ isophotal magnitude.  The same method was used on the
ground--based data to extend the stellar sample to areas beyond the ACS
coverage.

\subsection{KPNO 4m + MOSAIC $U$--band imaging}

Deep $U$--band images of the GOODS HDF--N field were obtained using the prime
focus MOSAIC camera on the KPNO Mayall 4--m telescope in March 2002.  A total
of 27.5 hours of exposure time were obtained in photometric conditions with
mean seeing FWHM~$= 1\farcs15$.  We reduced the data using the IRAF {\tt
mscred} package, which carries out bias subtraction, flat--fielding,
correction for geometric distortion, removal of amplifier cross--talk,
subtraction of an additive ``pupil ghost'', image registration, and
combination.

\subsection{ESO--MPI 2.2--m + WFI imaging}

An area of $\sim 0.4$ sq. degree around the CDF-S was surveyed with WFI on the
ESO--MPI 2.2--m telescope using the $U^{\prime}UBVRI$ passbands.  Much of
these data were obtained as part of the EIS, and are described in Arnouts et
al.\ 2001.  Additional $BVR$ images were obtained for the GOODS program and in
the course of the COMBO--17 project (Wolf et al.\ 2001).  The EIS, GOODS, and
COMBO data were reduced in a uniform fashion using the GaBoDs WFI reduction
pipeline (Schirmer et al.\ 2003).  We carefully checked the accuracy of the
photometric calibration using the color loci of stars.  These were compared to
synthetic color-color diagrams generated using the Gunn \& Stryker (1983)
spectrophotometric library, convolved with the instrumental passbands.
Zeropoints were adjusted as necessary, and are accurate to $< 0.1$ mag.

\subsection{ESO VLT + FORS1 imaging}

Additional, deep $R$ and $I$-band images were obtained with the 
ANTU telescope (UT1) at the VLT, using the FORS1 instrument. The
data and their reduction are described in Tozzi et al.\ (2001).  
The FORS1 images cover only a portion of the GOODS CDF--S field using 
several pointings with variable exposure time.   A portion
of the field is covered by relatively short $R$--band exposures.
The core region ($13\farcm6 \times 13\farcm6$) is quite deep, however,
and in particular is much deeper than the WFI imaging in the $I$--band.
The photometric zeropoints were checked and adjusted by comparing photometry 
for stars and galaxies in the WFI and FORS1 images, accounting for the 
passband differences, and agree on average
to $< 0.05$ mag.

\subsection{ESO NTT + SOFI imaging}

Near--infrared data in the $J$ and $K_s$ bands for the CDF--S were 
obtained with SOFI on the NTT as part of the EIS.  The observations 
and reductions are described in Vandame et al.\ 2001.  They consist 
of a 4$\times$4 grid of pointings covering 0.1 sq. deg.  SOFI $H$-band 
data were obtained and reduced by Moy et al.\ 2002, and cover 
a larger area.  Photometric zero-points were checked by comparing 
photometry of stars to measurements from the Two-Micron All-Sky 
Survey (2MASS).  

\subsection{ESO VLT + ISAAC imaging}

The GOODS program is gathering a mosaic of very deep $JHK_s$ images 
with ISAAC on the VLT as part of an ESO Large Programme.  Ultimately,
32 pointings will be used to cover the GOODS CDF--S field.  Data 
for 8 fields were obtained in 2001--2 and combined with ISAAC data 
from an ESO visitor program, kindly provided by E.\ Giallongo.  These
images cover only $\sim 50$~arcmin$^2$, but are much deeper than the 
SOFI data, with excellent image quality (mean FWHM~$\approx 0\farcs45$).
They were reduced with the EIS pipeline.  Their photometric zeropoints 
were adjusted to a common scale by comparison to objects in the SOFI 
images, and agree within 0.07 mag.  

\section{Source Catalogs}

Several different source catalogs for the ACS and ground--based
images are used in the accompanying papers.  In all cases, source 
identification and photometry were performed using SExtractor 
(Bertin \& Arnouts 1996).  Simulations were used to guide the choice
of detection thresholds and the size and shape of the convolution 
kernel to optimize the detection of faint galaxies while keeping 
spurious sources to a minimum.  Photometric uncertainties within
any aperture are computed using the normalized noise maps that were 
produced as part of the data reduction process.  However, 
some projects use detailed simulations with artificial sources
to assess the true photometric errors.  

For the ACS catalog, sources were detected in the $z_{850}$
mosaic, and photometry was carried out through matched apertures 
in the other ACS bands.  Several different ground--based catalogs 
were generated for the CDF--S, with detections based on the WFI 
$R$--band, SOFI $K$--band, and ISAAC $K$ and $J$--bands.  In each 
case, matched aperture photometry was measured for all of the 
other CDF--S ground--based images.  A wide--field HDF--N catalog,
based on Subaru $BVRIz$ imaging and the GOODS KPNO $U$ data, 
is presented by Capak et al.\ 2003, but is not used in the 
present set of Letters.  Where needed, ACS--to--ground--based 
colors were measured using images degraded to match the 
ground--based PSFs.   These provide optical--IR colors of ACS 
sources, as well as $U-B_{435}$ colors for Lyman break color 
selection.

Photometric redshifts for galaxies in the CDF--S (Mobasher et al.\ 2003)
were estimated using all of the available photometry ($U$ through $K_s$ 
bands) and the BPZ software of Ben\'{i}tez (2000).  They were tested 
against spectroscopic redshifts from the K20 survey (Cimatti et al.\ 2002), 
kindly provided by A.\ Cimatti, and also against redshifts measured 
with VLT+FORS2 for the GOODS program.  

\section{Sensitivity}

The sensitivity of the images depends on the sizes and shapes of the objects
of interest.  To guide the readers of these Letters, Table 2 provides two
sensitivity estimates (point source flux and limiting surface brightness) for
each of the GOODS space-- and ground--based imaging data sets.  These should
be taken as representative values only, since the actual depths of the imaging
data sets vary over the field of view, and depend on the details of source
size, PSF shape, crowding, etc.

Many of the papers in this issue rely on SExtractor detections in the ACS
\wz-band image. To assess the completeness limits for this catalog, we have
iteratively inserted sources into the image and re-run SExtractor. This is the
standard procedure used for point-source photometry to assess completeness and
photometric errors in color-magnitude diagrams. The problem has more
dimensions for galaxy photometry because the detectability of a galaxy depends
on the size and surface brightness profile. Our simulations uniformly populate
the magnitude range $20 < z_{850} < 28$ and the range of galaxy half-light
radii $0\farcs01 < r < 1\farcs5$. The input galaxies are a 50/50 mix of
spheroids --- galaxies with $r^{1/4}$-law surface-brightness profiles --- and
disks with exponential surface-brightness profiles. The spheroids are assumed
to be oblate and optically thin with an intrinsic axial ratio distribution
that is uniform in the range $0.3 < b/a < 0.9$.  The disks are modeled as
optically-thin oblate spheroids with a (very flat) intrinsic axial ratio of
$b/a = 0.05$.  Galaxies are viewed from random inclinations.  The simulated
galaxies are convolved with the PSF and inserted into the $z$-band image
without additional Poisson noise. For bright sources, the noise is thus
slightly underestimated, but for faint sources, the sky background completely
dominates anyway so this shortcut does not affect the results. The resulting
completeness limits in a plane of magnitude and half-light radius are shown in
Fig. 4.


\section{Summary}

The intent of this paper has been to provide a brief overview of the GOODS
project and to describe the data used to derive the scientific results
discussed elsewhere in this issue. The GOODS web site, which can be accessed
from http://www.stsci.edu/ftp/science/goods/, provides further details of the
project and access to the GOODS data.

\acknowledgments

Support for the GOODS {\it HST} Treasury program was provided by NASA through
grants HST-GO09425.01-A and HST-GO-09583.01. Additional support for this work,
which is also part of the {\it Space Infrared Telescope Facility (SIRTF)}
Legacy Science Program, was provided by NASA through Contract Number 1224666
issued by the Jet Propulsion Laboratory, California Institute of Technology
under NASA contract 1407. PM acknowledges support by NASA through grant
NAG5-11513.

A project of the scale and scope of GOODS could not succeed without the
dedicated efforts of a large number of people. We wish to thank Beth
Perriello, Bill Workman, Ian Jordan, Denise Taylor and David Soderblom for
their efforts in planning and scheduling the GOODS ACS observations. Guido De
Marchi provided expert assistance on ACS issues. Dorothy Fraquelli and the
OPUS group at STScI were instrumental in enabling the rapid turnaround of the
data for the supernova search. Mark Calvin, Matt Divens, and the computer
support group went well beyond the call of duty in their support of the
data-reduction computers, on several occasions rescuing the supernova searches
from impending disaster. We acknowledge the contributions of Len Cowie to the
GOODS project, and thank Peter Capak for providing the Subaru optical catalog
in advance of publication and helping to support the HDF--N supernova searches.
We thank Emmanuel Moy, Dimitra Rigopoulou and Pauline Barmby for providing the
SOFI $H$-images prior to publication, Emanuele Giallongo for contributing his
ISAAC imaging of the CDF--S to the GOODS public--release data set, and Andrea
Cimatti and the K20 team for providing spectroscopic redshifts that were used
to validate our photometric redshift estimates.

\clearpage

\begin{deluxetable}{llrcl}
\tabletypesize{\scriptsize}
\tablewidth{0pc}
\tablecaption{The Data}
\tablehead{
\colhead{Facility} & 
\colhead{Passbands} & 
\colhead{Area coverage\tablenotemark{a}} & 
\colhead{Angular resolution\tablenotemark{b}} & 
\colhead{Field} }
\startdata
\hst\ + ACS       & $BViz$    &  320\tablenotemark{c} & 0.125\tablenotemark{d}    & HDF-N + CDF-S \\
\cxo + ACIS       & 0.5-8 keV &  450\tablenotemark{e} & 0.85--10\tablenotemark{f} & HDF-N \\
\cxo + ACIS       & 0.5-8 keV &  390\tablenotemark{e} & 0.85-10\tablenotemark{f}                   & CDF-S \\
KPNO 4-m + MOSAIC & $U$       & 1800                  & 1.15                      & HDF-N \\
ESO 2.2-m + WFI   & $U'UBVRI$ & $>$1350               & 0.85--1.05                & CDF-S \\
ESO VLT + FORS1   & $RI$      &  175\tablenotemark{g} & 0.6--0.8                  & CDF-S \\
ESO NTT + SOFI    & $JK_s$    &  360                  & 0.65-1.05                 & CDF-S \\
ESO NTT + SOFI    & $H$       &  630                  & 0.55-0.85                 & CDF-S \\
ESO VLT + ISAAC   & $JHK_s$   &   50                  & 0.40-0.65                 & CDF-S \\
\enddata
\tablenotetext{a}{Total area covered, in arcmin$^2$.}
\tablenotetext{b}{PSF FWHM, in arcseconds}
\tablenotetext{c}{Area with 4--band ACS coverage.  The total area with 
$Viz$--band coverage is 365~arcmin$^2$.}
\tablenotetext{d}{Modal PSF FWHM of current version of drizzled image mosaics.
The intrinsic image quality of \hst/ACS is better;  future rereductions will
improve the net image quality.}
\tablenotetext{e}{Total \cxo\ coverage.  Exposure time and PSF, and hence
net sensitivity, are a strong function of area.}
\tablenotetext{f}{\cxo\ PSF quoted as 50\% encircled energy diameter, 
from on--axis to the maximum off--axis angle within the GOODS ACS area.}
\tablenotetext{g}{Deep area only.  The FORS $R$ mosaic covers an additional 
160\arcmin\ at shallower depth.}
\end{deluxetable}
\newpage

\begin{deluxetable}{lcccccccccc}
\tabletypesize{\scriptsize}
\tablewidth{0pc}
\tablecaption{Sensitivity for Current GOODS Optical and Near--Infrared Data}
\tablehead{
\colhead{Facility} & 
\colhead{$U^\prime$}& 
\colhead{$U$} & 
\colhead{$B$} & 
\colhead{$V$} & 
\colhead{$R$} & 
\colhead{$I$} & 
\colhead{$z$} & 
\colhead{$J$} & 
\colhead{$H$} & 
\colhead{$K_s$}}
\startdata
\hst+ACS    & \ndat & \ndat & 27.8  & 27.8  & \ndat & 27.1  & 26.6  & \ndat & \ndat & \ndat \\
\hst+ACS    & \ndat & \ndat & 28.4  & 28.4  & \ndat & 27.7  & 27.3  & \ndat & \ndat & \ndat \\

4--m MOSAIC & \ndat & 25.9  & \ndat & \ndat & \ndat & \ndat & \ndat & \ndat & \ndat & \ndat \\
4--m MOSAIC & \ndat & 29.0  & \ndat & \ndat & \ndat & \ndat & \ndat & \ndat & \ndat & \ndat \\

2.2--m WFI  & 25.0\tablenotemark{a}  & 24.7\tablenotemark{a}  
			    & 26.2  & 25.8  & 25.8  & 23.5  & \ndat & \ndat & \ndat & \ndat \\
2.2--m WFI  & 28.1\tablenotemark{a}  & 27.9\tablenotemark{a}  
			    & 29.3  & 28.9  & 28.9  & 26.6  & \ndat & \ndat & \ndat & \ndat \\

VLT FORS1   & \ndat & \ndat & \ndat & \ndat & 26.2  & 25.9  & \ndat & \ndat & \ndat & \ndat \\
VLT FORS1   & \ndat & \ndat & \ndat & \ndat & 29.4  & 29.0  & \ndat & \ndat & \ndat & \ndat \\

NTT SOFI    & \ndat & \ndat & \ndat & \ndat & \ndat & \ndat & \ndat & 22.8  & 22.0  & 21.8  \\
NTT SOFI    & \ndat & \ndat & \ndat & \ndat & \ndat & \ndat & \ndat & 25.9  & 25.1  & 25.0  \\

VLT ISAAC   & \ndat & \ndat & \ndat & \ndat & \ndat & \ndat & \ndat & 25.5  & 24.9  & 25.1  \\
VLT ISAAC   & \ndat & \ndat & \ndat & \ndat & \ndat & \ndat & \ndat & 27.9  & 27.3  & 27.4  \\
\enddata

\tablenotetext{\ } {For each telescope + instrument combination, the first line gives the 
10$\sigma$ point--source sensitivity within an aperture diameter of 0\farcs2 for \hst, 1\farcs0 
for ISAAC, and 2\farcs0 for other ground--based data.  The second line gives the $1\sigma$ surface 
brightness fluctuations in an aperture with area 1 arcsec$^2$.  The values reported are medians
over the area covered by the \hst/ACS imaging, except for the FORS1 $R$ data which is a value
for the area covered by the deeper portion of the data.  
Units in AB mag (Oke 1974) and AB mag arsec$^{-2}$, respectively.} 

\tablenotetext{a}{The WFI $U$--band is highly non--standard (see Arnouts et al.\ 2001).
The so--called $U^\prime$ filter for WFI is closer to a standard $U$ passband.}


\end{deluxetable}

\clearpage

%
%
%
%
%
 

\clearpage

\begin{figure}
\figurenum{1}
\epsscale{0.8}
\plotone{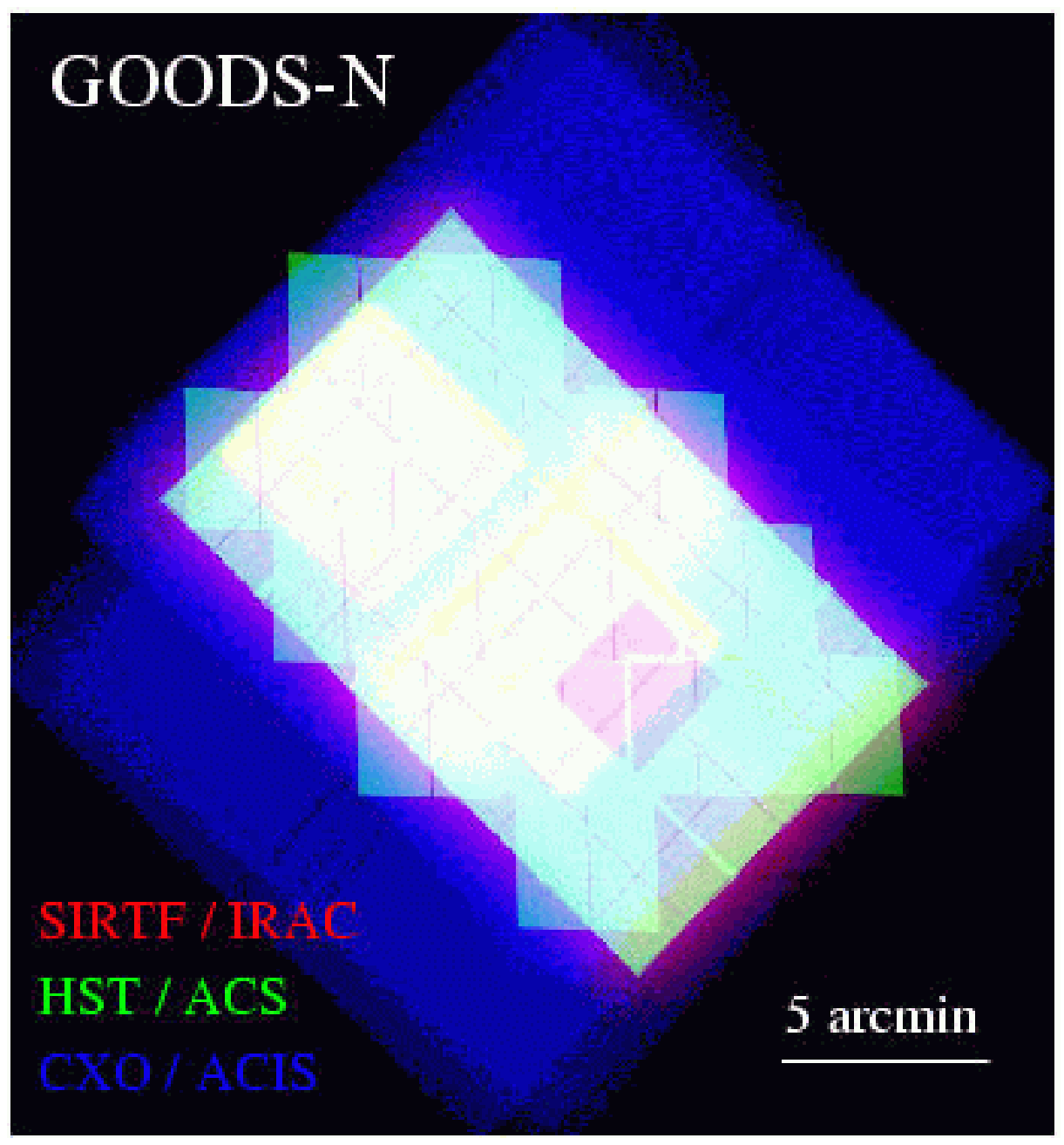}
\caption{\figonecap}
\end{figure}
\newpage

\begin{figure}
\figurenum{2}
\epsscale{1.0}
\plotone{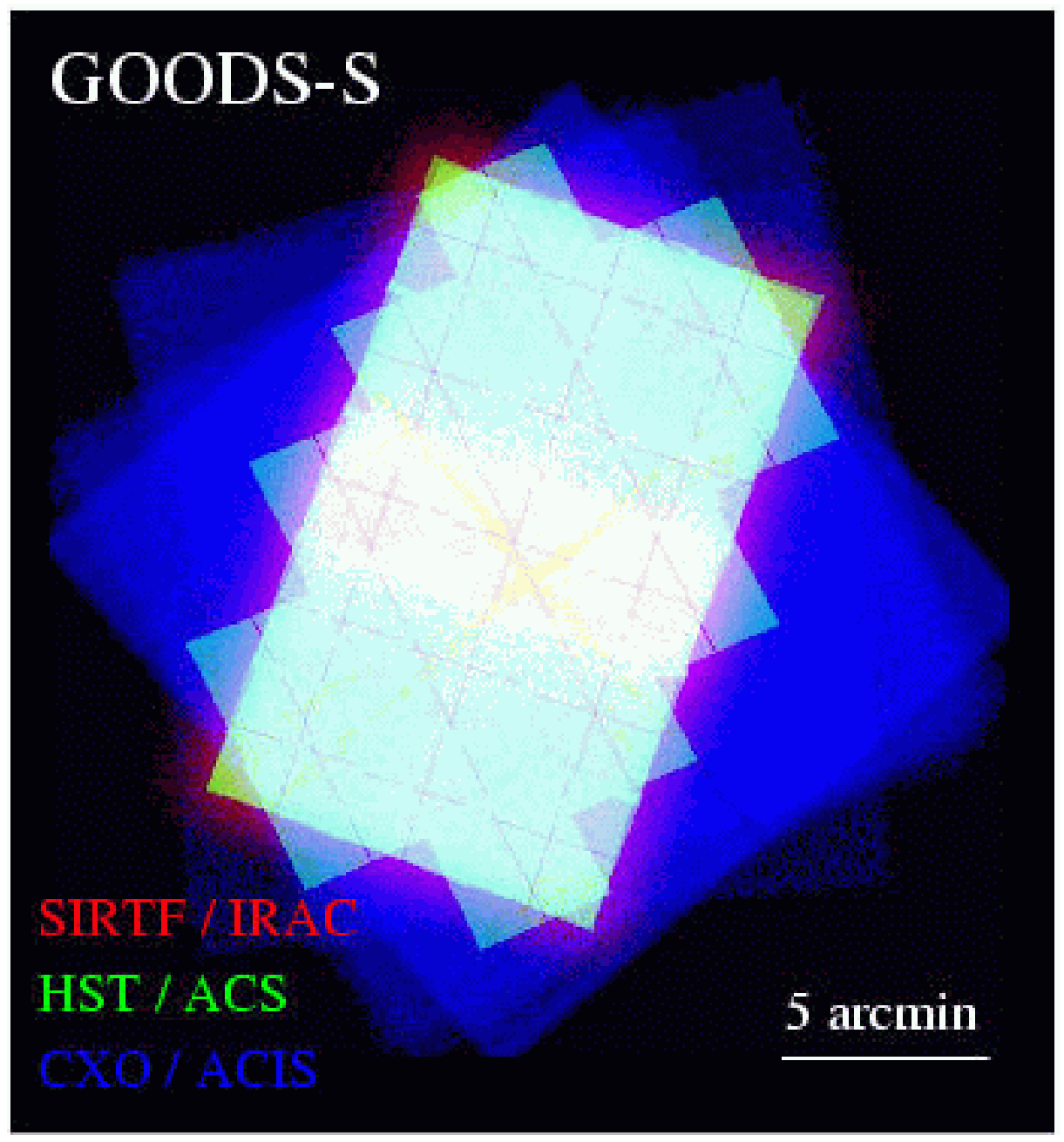}
\caption{\figtwocap}
\end{figure}
\newpage

\begin{figure}
\figurenum{3}
\epsscale{1.0}
\plotone{f3.eps}
\caption{\figthreecap}
\end{figure}
\newpage
 
\begin{figure}
\figurenum{4}
\epsscale{1.0}
\plotone{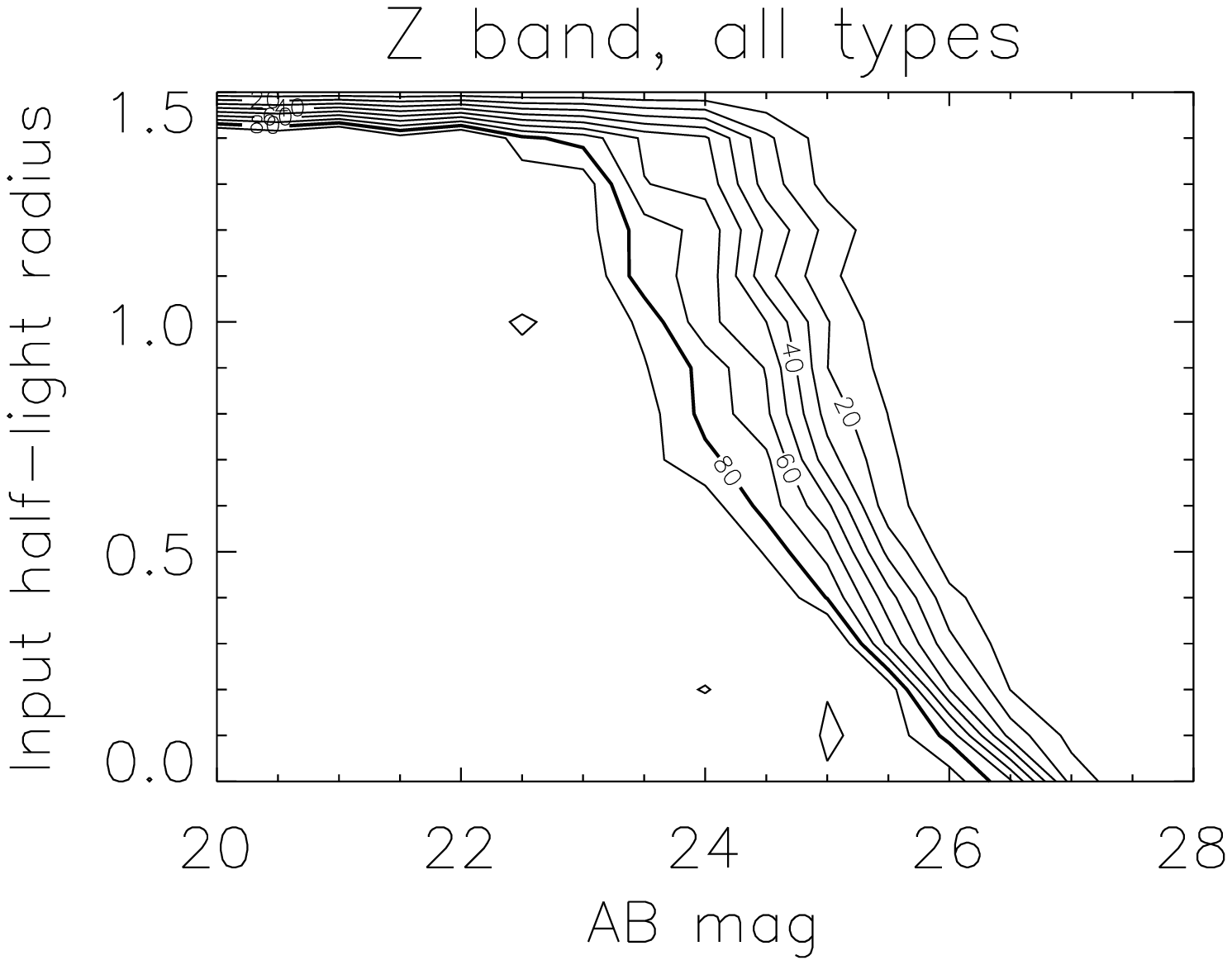}
\caption{\figfourcap}
\end{figure}
\newpage


\begin{references}

\reference{} Alexander, D., et al., 2003, AJ, in press, astro-ph/0304392
\reference{} Arnouts, S., et al., 2001, A\&A, 379, 740
\reference{} Ben\'{i}tez, N., 2000, ApJ, 536, 571
\reference{} Bertin, E., 2002, SWarp User's Guide
\reference{} Bertin, E., and Arnouts, S., 1996, A\&A, 117, 393
\reference{} Bunker, A., Smith, J., Spinrad, H., Stern, D. \&
	     Warren, S., 2003, astro-ph/0303290.
\reference{} Capak, P., et al., 2003, ApJ, submitted
\reference{} Cimatti, A., et al., 2002, A\&A, 392, 395
\reference{} Cohen, J., Hogg, D. W., Blandford, R., Cowie, L. L., 
	     Hu, E. Songaila, A., Shopbell, P. \& Richberg, K., 2000, 
	     ApJ, 538, 29
\reference{} Dawson, S., McCrady, N., Stern, D., Eckart, M., Spinrad, H.,
	     Liu, M., \& Graham, J. R., 2002, astro-ph/0212240 
\reference{} Dickinson, M., et al., 2003, this volume.
\reference{} Ferguson, H. C., Dickinson, M., \& Williams, 2000, ARAA, 38, 667
\reference{} Fruchter, A. S. \& Hook, R. N. 2002, PASP 114, 144 
\reference{} Giacconi, R. et al., 2002, ApJS, 139, 369 
\reference{} Gunn, J., \& Stryker, 1983, ApJS, 52, 121
\reference{} Koekemoer, A. M., Fruchter, A. S. Hack, W. \& Hook, R. N. 2003,
             {\it HST} Calibration Workshop (STScI: Baltimore)
\reference{} Madau, P., Pozzetti, L., \& Dickinson, M. E. 1998, ApJ, 498, 106
\reference{} Mobasher, B., et al., this volume
\reference{} Moy E., et al. 2002, astro-ph/0211247
\reference{} Oke, J. B. 1974, ApJS, 27, 21
\reference{} Pavlovsky, C., et al. 2002, ACS Instrument
             Handbook, Version 3.0, (Baltimore: STScI). 
\reference{} Riess, A., et al., 2003, this volume
\reference{} Roche, N., Dunlop, J. \& Almaini, O., 2003, astro-ph/0303206
\reference{} Schirmer, M., Erben, T., Schneider, P., Pietrzynski, G., Gieren, W., 
             Micol, A., Pierfederici, F. 2003, astro-ph/0305172
\reference{} Somerville, R. S., et al., 2003, this volume
\reference{} Spergel, D., et al., 2003, ApJ, submitted
\reference{} Stanway, E., Bunker, A., \& McMahon, R., 2003, astro-ph/0302212
\reference{} Steidel, C. C., Giavalisco, M., Pettini, M., Dickinson, M., \& 
             Adelberger, K. 1996, ApJ, 
\reference{} Tozzi, P., et al., 2001, ApJ, 562, 42
\reference{} Vandame, B., et al., 2001, A\&A, submitted (astro-ph/0102300), 
\reference{} Williams, R. E. et al. 1996, AJ, 112, 1335
\reference{} Williams, R. E. et al. 2000, AJ, 120, 2735
\reference{} Wolf, C., Dye, S., Kleinheinrich, M., Meisenheimer, K., Rix, H.-W., 
		\& Wisotzki, L., 2001, A\&A, 377, 442
\reference{} Yan, H., Windhorst, R. A., R\"ottgering, H. J., Cohen, S. H.,
	     Odewhan, S. C., Chapman, S. C. \& Keel, W. C., 2003, ApJ, 585, 67 


\end{references}
\end{document}